\documentclass[shortnote,twocolumn]{jpsj3}
\usepackage{color,graphicx}
\usepackage{amsmath,amssymb}
\usepackage{braket}
\usepackage{subfigure}

\graphicspath{{fig/}}

\title{Fair Sampling by Simulated Annealing on Quantum Annealer}

\author{Masayuki Yamamoto$^1$\thanks{m.yamamoto@dc.tohoku.ac.jp}, Masayuki Ohzeki$^{1,2}$, and Kazuyuki Tanaka$^1$}
\inst{$^1$Graduate School of Information Sciences, Tohoku University, Sendai 980-8579, Japan \\
$^2$Institute of Innovative Research, Tokyo Institute of Technology, Kanagawa, 226-8503, Japan}

\abst{Conventional quantum annealing does not sample all ground states fairly. 
We demonstrate that fair sampling can be achieved by performing simulated annealing on a quantum annealer. 
We discuss the problems that occur when implementing this method and propose an alternative way to overcome them. 
We numerically verify the fair sampling ability of our method in a small-scale toy model.}

\begin{document}
\maketitle

Quantum annealing (QA) \cite{PhysRevE.58.5355,Brooke779,Farhi472,Santoro2427,Santoro_2006,RevModPhys.80.1061,doi:10.1063/1.2995837} is a metaheuristic for solving combinatorial optimization problems, in which the optimal solution(s) of a combinatorial optimization problem is represented as the ground state(s) of a certain Hamiltonian. 
Several important applications that have multiple solutions rely on the \textit{fair sampling} ability of an algorithm---i.e., the ability to sample all the ground states of a degenerate problem with equal probability---, such as satisfiability filters \cite{weaver2012satisfiability,10.1007/978-3-319-24318-4_9,SciPostPhys.2.2.013} or machine learning \cite{doi:10.1162/089976602760128018,eslami2014shape}. 
However, it was demonstrated both theoretically and experimentally that QA under transverse-field driving lacks the fair sampling ability \cite{Matsuda_2009,PhysRevLett.118.070502}. Although QA with more complex drivers was investigated in \citen{PhysRevA.100.030303}, it was concluded that a considerably complex driver is required to achieve fair sampling.

We propose a different approach to fair sampling using the method to perform simulated annealing (SA) on QA proposed by Somma, Batista, and Ortiz \cite{PhysRevLett.99.030603} (hereafter referred to the SBO method). 
In SA, the system maintains thermal equilibrium by lowering the temperature quasistatically. 
In the Boltzmann distribution, which is in thermal equilibrium, the probabilities become equal if the energies become equal; hence, it is considered that SA can achieve fair sampling \cite{4767596}. 
In this present paper, we propose a fair sampling method using the SBO method and discuss the problems that occur when implementing it.

We define the cost function (problem Hamiltonian) as
\begin{align}
H_0(\boldsymbol{\sigma})=-\sum_{i<j}J_{ij}\sigma_i\sigma_j-\sum_{i=1}^{N}h_i\sigma_i,
\end{align}
where $\sigma_i$ is the $i$-th binary variable taking $\pm 1$, and $N$ is the total number of variables. 
The statistical mechanical expectation value at a temperature $T$ of a physical quantity $A(\boldsymbol{\sigma})$ is expressed as
\begin{align}
\langle A\rangle_T=\frac{1}{Z(T)}\sum_{\sigma}e^{-\beta H_0(\boldsymbol{\sigma})}A(\boldsymbol{\sigma}),
\end{align}
where $Z(T)=\sum_{\boldsymbol{\sigma}}e^{-\beta H_0(\boldsymbol{\sigma})}$ and $\beta=1/T$. 
We define the quantum state as
\begin{align}
\ket{\psi(T)}=\frac{1}{\sqrt{Z(T)}}e^{-\frac{1}{2}\beta\hat{H}_0}\sum_{\boldsymbol{\sigma}}\ket{\boldsymbol{\sigma}},
\end{align}
then $\braket{\psi(T)|\hat{A}|\psi(T)}=\langle A\rangle_T$ holds true, where $\hat{\sigma}_i^z$ is the $z$ component of the Pauli operator acting on site $i$, $\hat{A}$ is the operator in which $\sigma_i$ in $A(\boldsymbol{\sigma})$ is replaced with $\hat{\sigma}_i^z$, and the same applies to $\hat{H}_0$. 
According to \citen{PhysRevLett.99.030603}, $\ket{\psi(T)}$ is the unique ground state of the Hamiltonian.
\begin{align}
\hat{H}_{\mathrm{S}}(T)=-\chi(T)\sum_{i=1}^{N}(\hat{\sigma}_i^x-e^{\beta\hat{H}_i}),
\end{align}
where $\hat{\sigma}_i^x$ is the $x$ component of the Pauli operator acting on site $i$, and $\hat{H}_i$ is the Hamiltonian containing only the terms in $\hat{H}_0$ that contain $\hat{\sigma}_i^z$, i.e. $\hat{H}_i=-\sum_{j=1}^{N}J_{ij}\hat{\sigma}_i^z\hat{\sigma}_j^z-h_i\hat{\sigma}_i^z$. $\chi(T)$ is defined as $\chi(T)=e^{-\beta p}$ ($p\sim \max_i\lvert\hat{H}_i\rvert$) such that $\hat{H}_\mathrm{S}(T)$ does not diverge at $T\to 0$.

The method of fair sampling under the total Hamiltonian $\hat{H}_\mathrm{S}(T)$ is as follows. First, prepare the state $\prod_{i=1}^{N}(\ket{\uparrow}_i+\ket{\downarrow}_i)/\sqrt{2}$, which is the ground state of $\hat{H}_\mathrm{S}(T\to\infty)$. 
Next, change the parameter $T$ adiabatically from $T=\infty$ to $T=0$ under the Hamiltonian $\hat{H}_\mathrm{S}(T)$. 
At each instance, the state maintains $\ket{\psi(T)}$, which is the ground state of $\hat{H}_\mathrm{S}(T)$ and corresponds to the Boltzmann distribution of temperature $T$. 
Finally, in $T\to 0$, $\ket{\psi(T\to 0}$ is the state in which the ground states of $\hat{H}_0$ are uniformly superposed. 
In this way, fair sampling is achieved.

When applying the SBO method to practical problems, a real quantum annealer such as a D-Wave machine is used; however, several problems arise. 
First, the temperature dependence of the interactions and the longitudinal fields differ for different places, as can be seen by expanding the second term of $\hat{H}_\mathrm{S}(T)$. 
In the current D-Wave machine, it is difficult for the interaction to change with time for varying positions. 
Second, the body interactions arising are generally more than two, as can be seen by expanding $e^{\beta\hat{H}_i}$. 
However, the current D-Wave machine can implement at most two body interactions. 
For these reasons, the application of the SBO method requires the expansion of machine functions.

Alternatively, we propose a slightly modified method that can avoid the problems on SBO one as explained above. 
We define the total Hamiltonian as
\begin{align}
\hat{H}(t)=\frac{t}{\tau}\hat{H}_{\mathrm{S}}(T)-\left(1-\frac{t}{\tau}\right)\sum_{i=1}^{N}\hat{\sigma}_i^x,
\end{align}
where $\tau$ is the annealing time. 
By setting the appropriate temperature parameter $T$ and performing QA under this Hamiltonian, we obtain the ground state of $\hat{H}_{\mathrm{S}}(T)$ at $t=\tau$, i.e., the state $\ket{\psi(T)}$, corresponding to the Boltzmann distribution at temperature $T$. Hereafter, this is referred to as the SBO+QA method, which enables the sampling of the Boltzmann distribution at any arbitrary temperature. 
In particular, fair sampling is achieved by setting the temperature $T$ sufficiently low. 
However, this method experiences a new problem, i.e., it is necessary to perform measurements in the finite strength of the transverse field because the final Hamiltonian $\hat{H}_{\mathrm{S}}(T)$ has a transverse-field term. 
The current D-Wave machine cannot be used for measurements in the presence of a transverse field. Instead, we may substitute a quantum quench such that the annealing schedule can be tuned. 
Note that the problem of more than two body interactions still exists.

\begin{figure}
	\centering
	\includegraphics[width=0.2\linewidth]{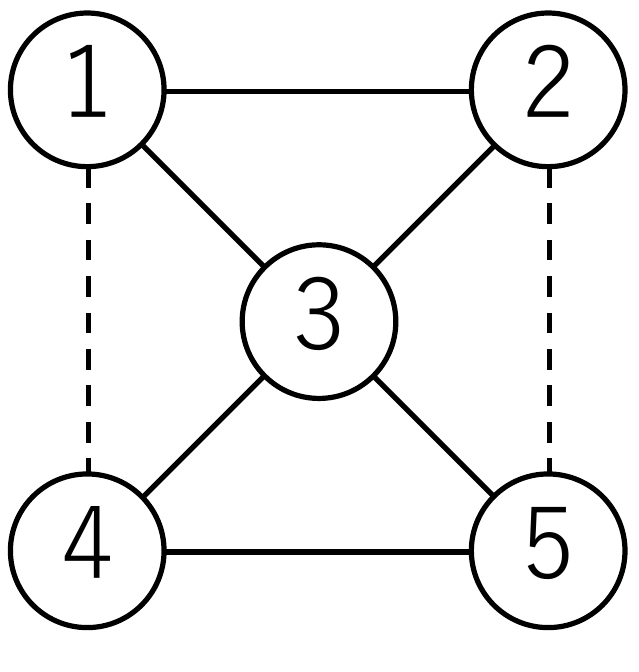}
	\caption{Five-spin toy model; the solid lines represent the ferromagnetic interactions ($J_{ij}=+1$) and the dashed lines represent the anti-ferromagnetic interactions ($J_{ij}=-1$).}
	\label{fig:problem}
\end{figure}

\begin{figure}
	\centering
	\subfigure{\includegraphics[width=0.49\linewidth]{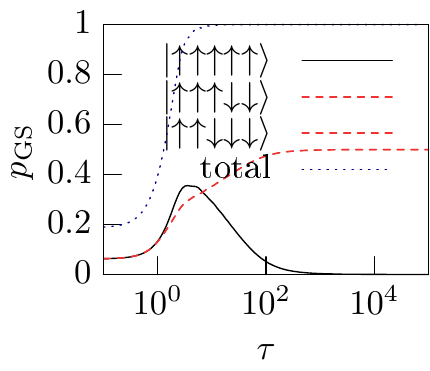}}
	\subfigure{\includegraphics[width=0.49\linewidth]{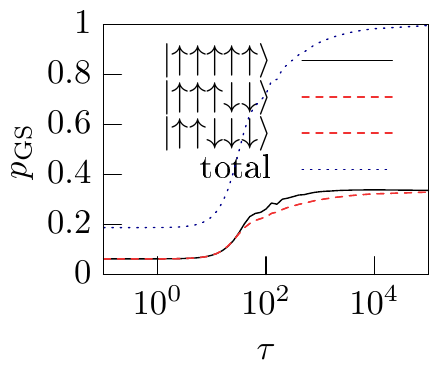}}
	\subfigure{\includegraphics[width=0.49\linewidth]{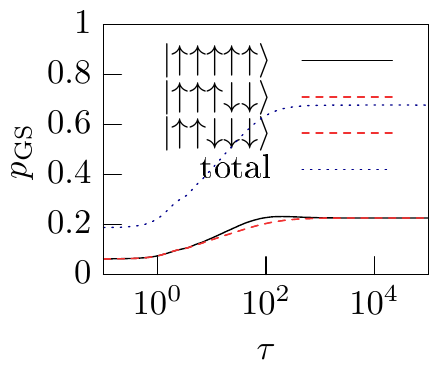}}
	\subfigure{\includegraphics[width=0.49\linewidth]{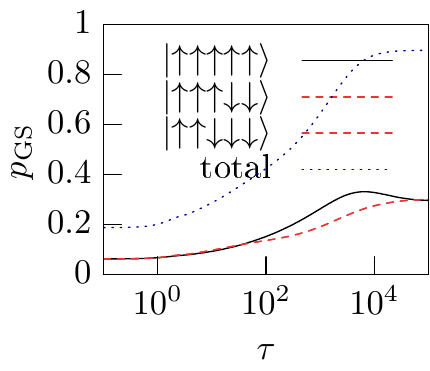}}
	\caption{Annealing time dependence of the probabilities when the ground states appear; upper left: results by QA, only two states are reached after a large $\tau$; upper right: the SBO method determines all the ground states fairly; bottom left and right: results by the SBO+QA method at $\beta=1$ and $2$, respectively. All the ground states emerge with equal probability and the excited states appear correspondingly. The total probability is higher when $\beta=2$ than $\beta=1$.}
	\label{fig:result}
\end{figure}

To verify the fair sampling abilities of the SBO and SBO+QA methods, we assume the problem shown in Fig. \ref{fig:problem} as a small-scale toy model \cite{Matsuda_2009,PhysRevA.100.030303}. 
This problem has six ground states: $\ket{\uparrow\uparrow\uparrow\uparrow\uparrow}$, $\ket{\uparrow\uparrow\uparrow\downarrow\downarrow}$, $\ket{\uparrow\uparrow\downarrow\downarrow\downarrow}$, and the states in which all the spins are inverted. 
As there is no longitudinal field, spin inversion symmetry exists, and hence, the probabilities of the two states in which the spins are inverted are equal. 
Therefore, we will focus on the above mentioned three states. 
We investigated the dependence of the probabilities $p_{\mathrm{GS}}$ such that the ground states appear on the annealing time $\tau$ in the conventional QA, SBO, and SBO+QA methods by numerically solving the Schr\"{o}dinger equation using QuTiP \cite{JOHANSSON20121760,JOHANSSON20131234}. 
For the SBO method, the temperature schedule was $\beta(t)=-10\ln(1-t/\tau)$. 
The SBO+QA method was examined for the case where $\beta=1$ and $2$. 
The results are shown in Fig. \ref{fig:result}. 
In conventional QA, the probability for $\ket{\uparrow\uparrow\uparrow\uparrow\uparrow}$ converges to 0, and those for $\ket{\uparrow\uparrow\uparrow\downarrow\downarrow}$ and $\ket{\uparrow\uparrow\downarrow\downarrow\downarrow}$ converge to $1/2$. 
The SBO method achieves fair sampling because all the probabilities have converged to $1/3$. For the SBO+QA method, all the ground states as well as the excited states emerge with equal probability owing to the finite temperature effect. 
The above results confirm that fair sampling has been achieved.

In this paper, we proposed a fair sampling method using the SBO method that implements SA on QA. We further proposed the SBO+QA method, which enables the sampling of the Boltzmann distributions at any temperature. 
Using the small-scale toy model, it was confirmed that these methods achieve fair sampling 
if the annealing time is sufficiently long. While using these methods on a real quantum annealer, such as a D-Wave machine, several enhancements are required. 
In addition, as real machines operate at finite temperatures, it is necessary to consider thermal effects that may appear. 
One way to theoretically estimate them is to use an open quantum system, such as the quantum adiabatic master equation \cite{Albash_2012} and more simply the interpolation between quantum dynamics and thermodynamics \cite{PhysRevA.97.022312,doi:10.7566/JPSJ.88.061008}. 
On resolving these issues, the methods we have proposed will enable several important applications.

\acknowledgment
The authors would like to thank Manaka Okuyama for the fruitful discussions. 
The present work was financially supported by JSPS KAKENHI Grant No. 18H03303 and 19H01095, and the JST-CREST (No.JPMJCR1402) for Japan Science and Technology Agency.

\end{document}